\documentclass[a4paper]{article}

\usepackage{INTERSPEECH2020}

\usepackage[utf8]{inputenc}
\usepackage{color}

\usepackage[unicode]{hyperref}
\hypersetup{bookmarksopen=true,bookmarksnumbered=true}

\hypersetup{
	colorlinks=true,
	linkcolor=black,
	citecolor=black,
	filecolor=black,
	urlcolor=black
}

\title{Ultrasound-based Articulatory-to-Acoustic Mapping \\ with WaveGlow Speech Synthesis}
\name{Tamás Gábor Csapó$^{1,2}$, Csaba Zainkó$^{1}$,
 László Tóth$^{3}$, Gábor Gosztolya$^{3,4}$, 
Alexandra Markó$^{2,5}$}
\address{
  $^1$Department of Telecommunications and Media Informatics, \\
	Budapest University of Technology and Economics, Budapest, Hungary \\
	$^2$MTA-ELTE Lendület Lingual Articulation Research Group, Budapest, Hungary \\
	$^3$Institute of Informatics, University of Szeged, Hungary\\
	$^4$MTA-SZTE Research Group on Artificial Intelligence, Szeged, Hungary \\
  $^5$Department of Applied Linguistics and Phonetics, Eötvös Loránd University, Budapest, Hungary
 }
\email{\{csapot, zainko\}@tmit.bme.hu, \{tothl, ggabor\}@inf.u-szeged.hu, marko.alexandra@btk.elte.hu}

\begin{document}

\maketitle
\begin{abstract}
For articulatory-to-acoustic mapping using deep neural networks, typically spectral and excitation parameters of vocoders have been used as the training targets. However, vocoding often results in buzzy and muffled final speech quality.
Therefore, in this paper on ultrasound-based articulatory-to-acoustic conversion, we use a flow-based neural vocoder (WaveGlow) pre-trained on a large amount of English and Hungarian speech data. The inputs of the convolutional neural network are ultrasound tongue images. The training target is the 80-dimensional mel-spectrogram, which results in a finer detailed spectral representation than the previously used 25-dimensional Mel-Generalized Cepstrum. From the output of the ultrasound-to-mel-spectrogram prediction, WaveGlow inference results in synthesized speech.
We compare the proposed WaveGlow-based system with a continuous vocoder which does not use strict voiced/unvoiced decision when predicting F0. The results demonstrate that during the articulatory-to-acoustic mapping experiments, the WaveGlow neural vocoder produces significantly more natural synthesized speech than the baseline system. Besides, the advantage of WaveGlow is that F0 is included in the mel-spectrogram representation, and it is not necessary to predict the excitation separately.

\end{abstract}

\section{Introduction}

Articulatory-to-acoustic mapping methods aim to synthesize speech signal directly from articulatory input, applying the theory that articulatory movements are directly linked with the acoustic speech signal in the speech production process. A recent potential application of this mapping is a “Silent Speech Interface” (SSI~\cite{Denby2010,Schultz2017a}), which has the main idea of recording the soundless articulatory movement, and automatically generating speech from the movement information, while the subject is not producing any sound. Such an SSI system can be highly useful for the speaking impaired (e.g. after laryngectomy or elderly people), and for scenarios where regular speech is not feasible, but information should be transmitted from the speaker (e.g. extremely noisy environments or military applications).

For the articulatory-to-acoustic mapping, the typical input can be electromagnetic articulography (EMA)~\cite{Cao2018,Taguchi2018}, ultrasound tongue imaging (UTI)~\cite{Denby2004,Hueber2010,Hueber2011,Jaumard-Hakoun2016,Tatulli2017,Csapo2017c,Grosz2018,Toth2018,Moliner2019,Gosztolya2019,Csapo2019}, permanent magnetic articulography (PMA)~\cite{Gonzalez2017a}, surface electromyography (sEMG)~\cite{Janke2017,Diener2018a}, Non-Audible Murmur (NAM)~\cite{Shah2018}, electro-optical stomatography~\cite{Stone2016} or video of the lip movements~\cite{Hueber2010,Ephrat2017,Sun2018}.
From another aspect, there are two distinct ways of SSI solutions, namely `direct synthesis' and `recognition-and-synthesis'~\cite{Schultz2017a}. In
the first case, the speech signal is generated without an intermediate step, directly from the articulatory data~\cite{Cao2018,Taguchi2018,Denby2004,Hueber2011,Jaumard-Hakoun2016,Csapo2017c,Grosz2018,Moliner2019,Gosztolya2019,Csapo2019,Gonzalez2017a,Janke2017,Diener2018a,Shah2018,Ephrat2017}. In the second case, silent speech recognition (SSR) is applied on the biosignal which extracts the content spoken by the person (i.e. the result of this step is text); this step is then followed by text-to-speech (TTS) synthesis~\cite{Hueber2010,Tatulli2017,Toth2018,Stone2016,Sun2018}. 
In the SSR+TTS approach, any information related to speech prosody is lost, whereas it may be kept with direct synthesis. Also, the smaller delay by the direct synthesis approach might enable conversational use.

For the direct conversion, typically, vocoders are used, which synthesize speech from the spectral parameters predicted by the DNNs from the articulatory input. One of the spectral representations that was found to be useful earlier for statistical parametric speech synthesis is Mel-Generalized Cepstrum in Line Spectral Pair form (MGC-LSP)~\cite{Csapo2015d,Csapo2016}. 

The early studies on articulatory-to-acoustic mapping typically applied a low-order spectral representation, for example, only 12 coefficients were used in \cite{Jaumard-Hakoun2016,Csapo2017c}.
Later, our team also experimented with using 22~kHz speech and 24-order MGC-LSP target~\cite{Csapo2019} (with the gain, having 25 dimensions altogether). Still, the 24-order MGC-LSP target is a relatively low-dimensional spectral representation, and this simple vocoder that we used in previous studies~\cite{Csapo2017c,Grosz2018,Toth2018,Moliner2019,Gosztolya2019,Csapo2019} can be a bottleneck in the ultrasound-to-speech mapping framework.

Besides the spectrum, the other aspect of direct conversion is to predict the source / excitation information, e.g.\ the fundamental frequency of speech. There have been only a few studies that attempted to predict the voicing feature and the F0 curve using ultrasound as input. Hueber et al.~experimented with predicting the V/UV parameter along with the spectral features of a vocoder, using ultrasound and lip video as input articulatory data~\cite{Hueber2011}. They applied a feed-forward deep neural network (DNN) for the V/UV prediction and achieved 
82\% accuracy.
In~\cite{Grosz2018}, we experimented with deep neural networks to perform articulatory-to-acoustic conversion from ultrasound frames (raw scanlines), with an emphasis on estimating the voicing feature and the F0 curve from the ultrasound input. We attained a correlation rate of 0.74 between the original and the predicted F0 curves, and an accuracy of 87\% in V/UV prediction (using the voice of only one speaker). However, in several cases, the inaccurate estimation of the voicing feature caused audible artefacts. Most recently, we used a continuous vocoder (which does not have a strict voiced/unvoiced decision~\cite{Csapo2015d,Csapo2016,Toth2016,Al-Radhi2017}), for the CNN-based UTI-to-F0 and UTI-to-MGC-LSP prediction~\cite{Csapo2019}. In the experiments with two male and two female speakers, we found that the continuous F0 was predicted with lower error, and the continuous vocoder produced similarly natural synthesized speech as the baseline vocoder using standard discontinuous F0. Also, the advantage of the improved vocoder is that, as all parameters are continuous, it is not necessary to train a separate network in classification mode for the voiced/unvoiced prediction.

\subsection{Neural vocoders}

Since the introduction of WaveNet in 2016~\cite{Oord2016}, neural vocoders are an exciting way of generating the raw samples of speech during text-to-speech synthesis (TTS). However, a problem with early WaveNet-like models was that they were computationally extremely expensive. Currently, state-of-the-art TTS models are based on parametric neural networks using improved versions of WaveNet-like neural vocoders. TTS synthesis is typically done in two steps: 1) the first step transforms the text into time-aligned features, such as a mel-spectrogram, 2) the second step transforms these spectral features to the speech signal. If we replace the first step (text-to-spectrogram) with articulation-to-spectrogram prediction, we can use the recent advances of the latter step directly for the purpose of articulation-to-speech synthesis.

One of the most recent types of neural vocoders, WaveGlow~\cite{Prenger2019} is a flow-based network capable of generating high-quality speech from mel-spectrograms. The advantage of the WaveGlow model is that it is relatively simple, yet the synthesis can be done faster than real-time. It can generate audio by sampling from a distribution (zero mean spherical Gaussian), conditioned on a mel-spectrogram.

\subsection{Contributions of this paper}

In this paper, we extend our earlier work on ultrasound-based articulatory-to-acoustic mapping. From the ultrasound tongue raw scanline input, we predict 80-dimensional STFT spectral representation, from which we synthesize speech with a WaveGlow model. We show that the use of a neural vocoder is advantageous compared to earlier vocoders, which applied source-filter separation.

\section{Methods}
\label{sec:methods}

\subsection{Articulatory data acquisition}

Two Hungarian male and two female subjects were recorded while reading sentences aloud (altogether 209 sentences each). The tongue movement was recorded in midsagittal orientation using the ``Micro'' ultrasound system of Articulate Instruments Ltd. at 81.67 fps. The speech signal was recorded with a Beyerdynamic TG H56c tan omnidirectional condenser microphone. The ultrasound data and the audio signals were synchronized using the tools provided by Articulate Instruments Ltd. In our experiments, the raw scanline data of the ultrasound was used as input of the networks, after being resized to 64$\times$128 pixels using bicubic interpolation. More details about the recording set-up and articulatory data can be found in~\cite{Csapo2017c}. The duration of the recordings was about 15 minutes, which was partitioned into training, validation and test sets in a 85-10-5 ratio.

\subsection{Speech data for neural vocoder}

WaveGlow~\cite{Prenger2019} provides a pretrained model trained on the LJSpeech database, from 24 hours of English audiobooks with a single female speaker. Our informal listening tests showed that the single speaker WaveGlow model can generate both male and female voice samples, but it performs weakly with low F0 values (which is typical for male speakers). However, since we have both male and female articulatory and acoustic recordings, we hypothesized that a multispeaker WaveGlow model will be more suitable for synthesizing speech. Therefore, we chose 5 male and 6 female Hungarian speakers (altogether 23k sentences, roughly 22 hours) from the PPSD database~\cite{Olaszy2013}. 

\subsection{Continuous vocoder (baseline)}

In the baseline vocoder, first, the speech recordings (digitized at 22~kHz) were analyzed using MGLSA~\cite{Imai1983} at a frame shift of 22\,050~Hz / 81.67~fps = 270 samples, which resulted in 24-order spectral (MGC-LSP) features \cite{Tokuda1994}. Next, continuous F0 (ContF0) is calculated on the input waveforms using the simple continuous pitch tracker~\cite{Garner2013}. After this step, Maximum Voiced Frequency (MVF) is calculated from the speech signal~\cite{Drugman2014b,Csapo2016}.
The continuous vocoder parameters (MGC-LSP, log-ContF0 and log-MVF) served as the training targets of the neural network in our speech synthesis experiments.

During the synthesis phase, voiced excitation is composed of residual excitation frames overlap-added pitch synchronously~\cite{Csapo2015d,Csapo2016,Al-Radhi2017}. This voiced excitation is lowpass filtered frame by frame at the frequency given by the MVF parameter. In the frequency range higher than the actual value of MVF, white noise is used. The voiced and unvoiced excitation is added together. Finally, an MGLSA filter is used to synthesize speech from the excitation and the MGC parameter stream~\cite{Imai1983}.

\subsection{WaveGlow neural vocoder}

During analysis, the mel-spectrogram was estimated from the Hungarian speech recordings (digitized at 22~kHz). Similarly to the original WaveGlow paper~\cite{Prenger2019}, 80 bins were used for mel-spectrogram using librosa mel-filter defaults (i.e.\ each bin is normalized by the filter length and the scale is the same as in HTK). FFT size and window size were both 1024 samples. For hop size, we chose 270 samples, in order to be in synchrony with the articulatory data.
This 80-dimensional mel-spectrogram served as the training target of the neural network.

NVIDIA provided a pretrained WaveGlow model using the LJSpeech database (WaveGlow-EN). Besides, another WaveGlow model was trained with the Hungarian data (WaveGlow-HU). This latter training was done on a server with eight V100 GPUs, altogether for 635k iterations.

In the synthesis phase, an interpolation in time was necessary, as the original WaveGlow models were trained with 22~kHz speech and 256 samples frame shift; for this we applied bicubic interpolation. Next, to smooth the predicted data, we used a Savitzky-Golay filter with a window size of five, and cubic interpolation. 
Finally, the synthesized speech is the result of the inference with the trained WaveGlow model (EN/HU) conditioned on the mel-spectrogram input~\cite{Prenger2019}.

\subsection{DNN training with the baseline vocoder}

Similarly to our previous study~\cite{Csapo2019}, here we used convolutional neural networks (CNN), but we further optimized manually the network structure and parameters. We trained speaker-specific CNN models using the training data (roughly 180~sentences). For each speaker, two neural networks were trained: one CNN for predicting the excitation features (log-ContF0 and log-MVF), and one for predicting the 25-dimensional MGC-LSP. All CNNs had one 64$\times$128 pixel ultrasound image as input, and had the same structure: two convolutional layers (kernel size: 13$\times$13, number of filters: 30 and 60), followed by max-pooling; and again two convolutional layers (filters: 90 and 120), followed by max-pooling. Finally, a fully connected layer was used with 1000 neurons. In all hidden layers, the Swish activation was used~\cite{Ramachandran2017}, and we applied dropout with 0.2 probability.
The cost function applied for the regression task was the mean-squared error (MSE). We used the SGD optimizer with manually chosen learning rate. We applied early stopping to avoid over-fitting: the network was trained for 100 epochs and was stopped when the validation loss did not decrease within 3 epochs.

\subsection{DNN training with the WaveGlow neural vocoder}

In the proposed system, one CNN is used for each speaker, with the same structure as for the baseline system: two convolutional layers, max-pooling, two convolutional layers, max-pooling, and a fully connected layer with 1000 neurons. The network had 64$\times$128 pixel images as input and was predicting the 80-dimensional mel-spectrogram features. The training procedures were the same as in the baseline setup.

\begin{figure}
\centering
\includegraphics[trim=1.0cm 0.0cm 1.5cm 1.0cm, clip=true, width=1.0\columnwidth]{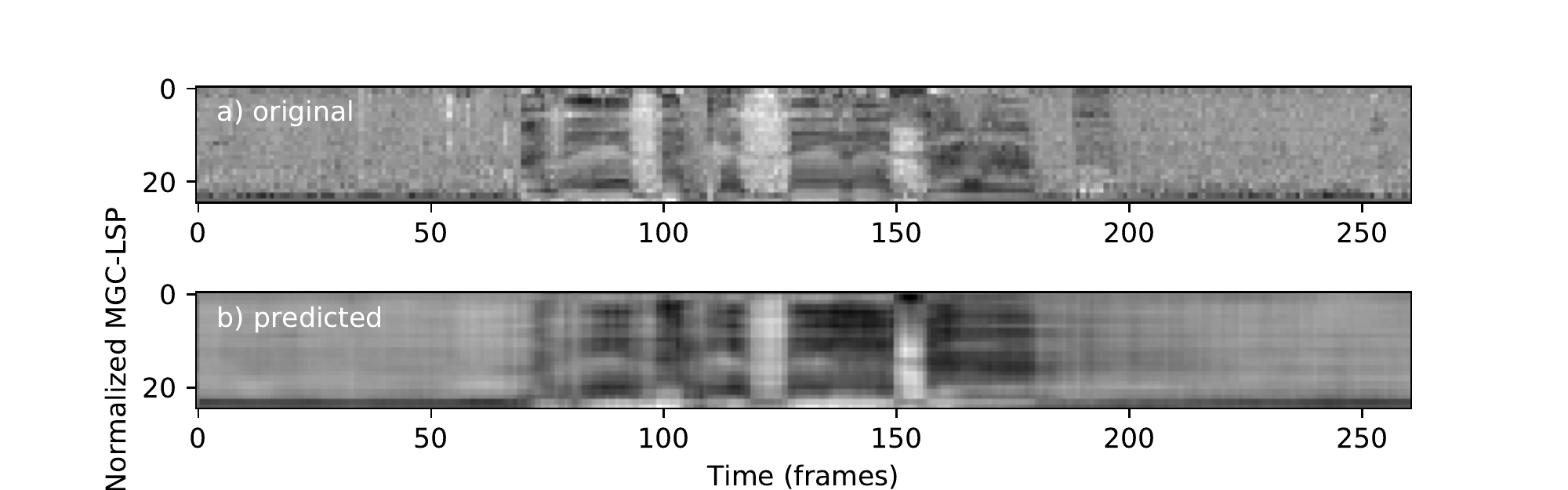}
\caption{Demonstration samples from a female speaker: normalized MGC-LSP using the baseline system.}
\label{fig:proposed_sample_MGC}
\includegraphics[trim=3.5cm 1.1cm 3.9cm 1.8cm, clip=true, width=1.0\columnwidth]{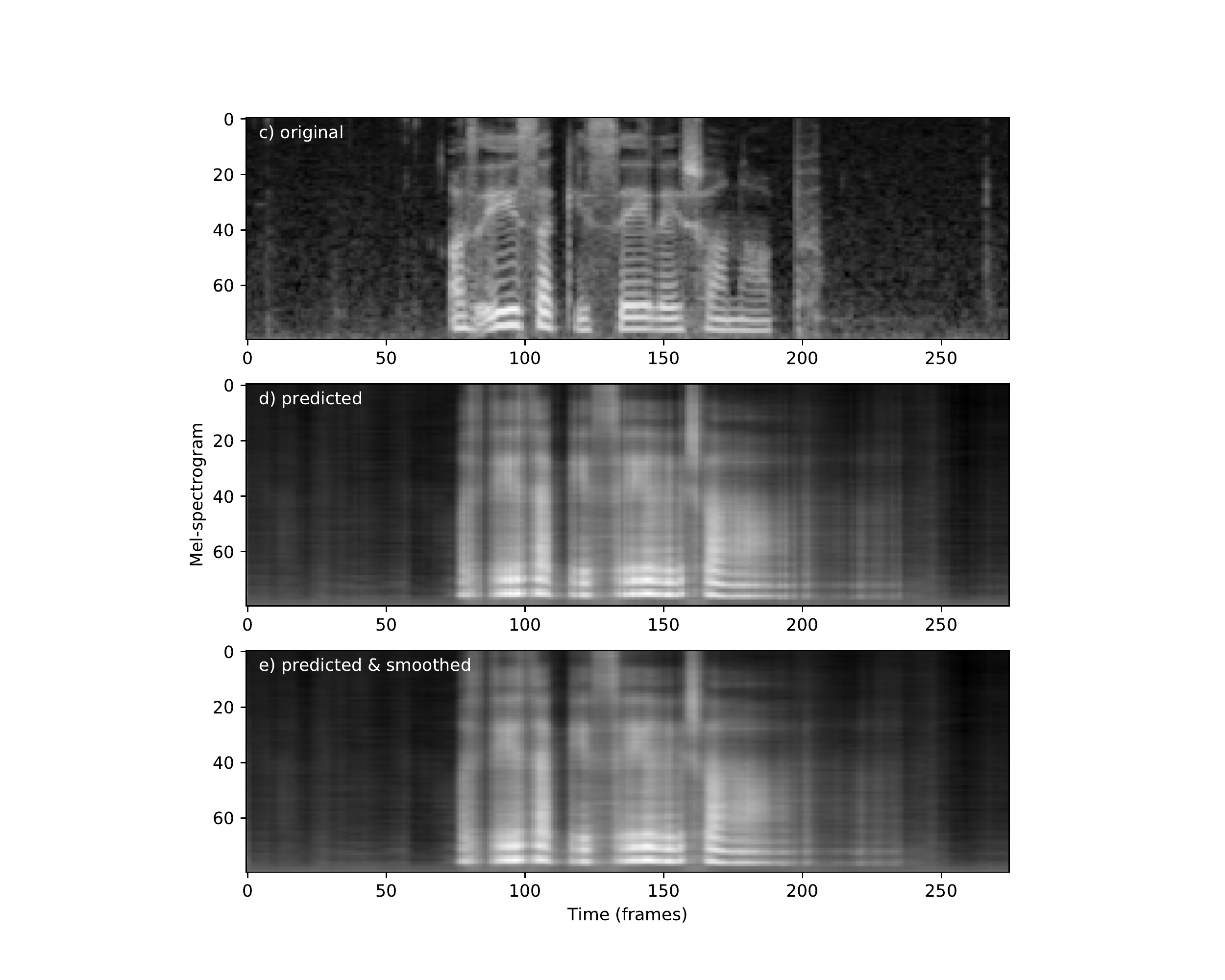}
\caption{Demonstration samples from a female speaker: normalized Mel-spectrogram using the proposed system.}
\label{fig:proposed_sample_Mel}
\end{figure}

\section{Experimental results}

\subsection{Demonstration sample}

A sample sentence (not being present in the training data) was chosen for demonstrating how the baseline and the proposed systems deal with the prediction of spectral parameters. Fig.~\ref{fig:proposed_sample_MGC} shows the spectral features of the baseline system (original and predicted MGC-LSP, normalized to zero mean and unit variance). In general, the generated speech starts at similar time as in the original recording, but it lasts longer -- probably as a result of inaudible post-speech tongue movement (around frames 190--220). Fig.~\ref{fig:proposed_sample_Mel} shows an example for the mel-spectogram prediction with the proposed system. The reason for the misaligned time scale is the different hop size (Fig.~\ref{fig:proposed_sample_MGC}: 270 samples, Fig.~\ref{fig:proposed_sample_Mel}: 256 samples). In the proposed system, the spectral details are similarly over-smoothed as in the baseline system, but the 80-dimensional mel-spectrogram contains more detailed information. Another difference between the baseline and the proposed systems is the way how they handle the voicing feature of speech excitation (i.e., in the proposed approach, the F0 information is included in the mel-spectrogram).

\begin{table} 
\caption{MCD scores on the test set.} \label{tab:objective_MCD}
\vspace{-2mm}
\centering
\begin{tabular}{l||c|c|c}
        & \multicolumn{3}{c}{{Mel-Cepstral Distortion (dB)}} \\
\cline{2-4}
 & ~~Continuous~~ & WaveGlow- & WaveGlow- \\
Speaker & ~~Vocoder~~ & EN & HU \\
\hline\hline
Speaker \#048 & 5.54 & 5.27 & 5.34   \\
Speaker \#049 & 5.67 & 5.66 & 5.65   \\
Speaker \#102 & 5.26 & 5.20 & 5.18   \\
Speaker \#103 & 5.41 & 5.34 & 5.37   \\
\hline
Mean & 5.47 & 5.37 & 5.38   \\
\end{tabular}
\vspace{-2mm}
\end{table}

\subsection{Objective evaluation}
\label{sec:objective}

After training the CNNs for each speaker and feature individually, we synthesized sentences, and measured objective differences. For this, Mean Square Error is not a suitable measure, as the (normalized) MGC-LSP features and the mel-spectrogram values have different scales; therefore, the MSE values by the baseline and proposed systems are not directly comparable.
Instead of MSE, the objective metric chosen in this test is the Mel-Cepstral Distortion (MCD,~\cite{Kubichek1993}). Lower MCD values indicate higher similarity. This metric not only evaluates the distance in the cepstral domain but also uses a perceptual scale in an effort to improve the accuracy of objective assessments, and is a standard way to evaluate text-to-speech synthesis systems. In general, the advantage of MCD is that it is better correlated with perceptual scores than other objective measures~\cite{Kubichek1993}. 

Table~\ref{tab:objective_MCD} shows the MCD results in dB for all the synthesized sentence types and speakers (lower values indicate higher spectral similarity). The best (lowest) MCD result, with an average value of 5.37~dB, was achieved when predicting mel-spectogram features and synthesizing with the English WaveGlow model. 
The Hungarian WaveGlow model scores second with an average MCD of 5.38~dB. Finally, the highest average error is attained by the baseline vocoder (MCD: 5.47~dB). By checking the MCD values speaker by speaker, we can see that both WaveGlow-EN and WaveGlow-HU were better than the baseline for all speakers, while there is some speaker dependency between the English and Hungarian versions (but these differences are mostly small in scale). 
Overall, according to the Mel-Ceptral Distortion measure, the proposed system is clearly better than the baseline vocoder.

\subsection{Subjective listening test}

In order to determine which proposed version is closer to natural speech, we conducted an online MUSHRA-like test~\cite{mushra}.

Our aim was to compare the natural sentences with the synthesized sentences of the baseline, the proposed approaches and a lower anchor system (the latter having constant F0 and predicted MGC-LSP).
In the test, the listeners had to rate the naturalness of each stimulus in a randomized order relative to the reference (which was the natural sentence), from 0 (very unnatural) to 100 (very natural). We chose four sentences from the test set of each speaker (altogether 16 sentences).
The variants appeared in randomized order (different for each listener).
The samples can be found at \url{http://smartlab.tmit.bme.hu/interspeech2020_UTI-to-STFT}.

\begin{figure}
\centering
\includegraphics[trim=0.25cm 0.3cm 0.2cm 0.25cm, clip=true, width=\columnwidth]{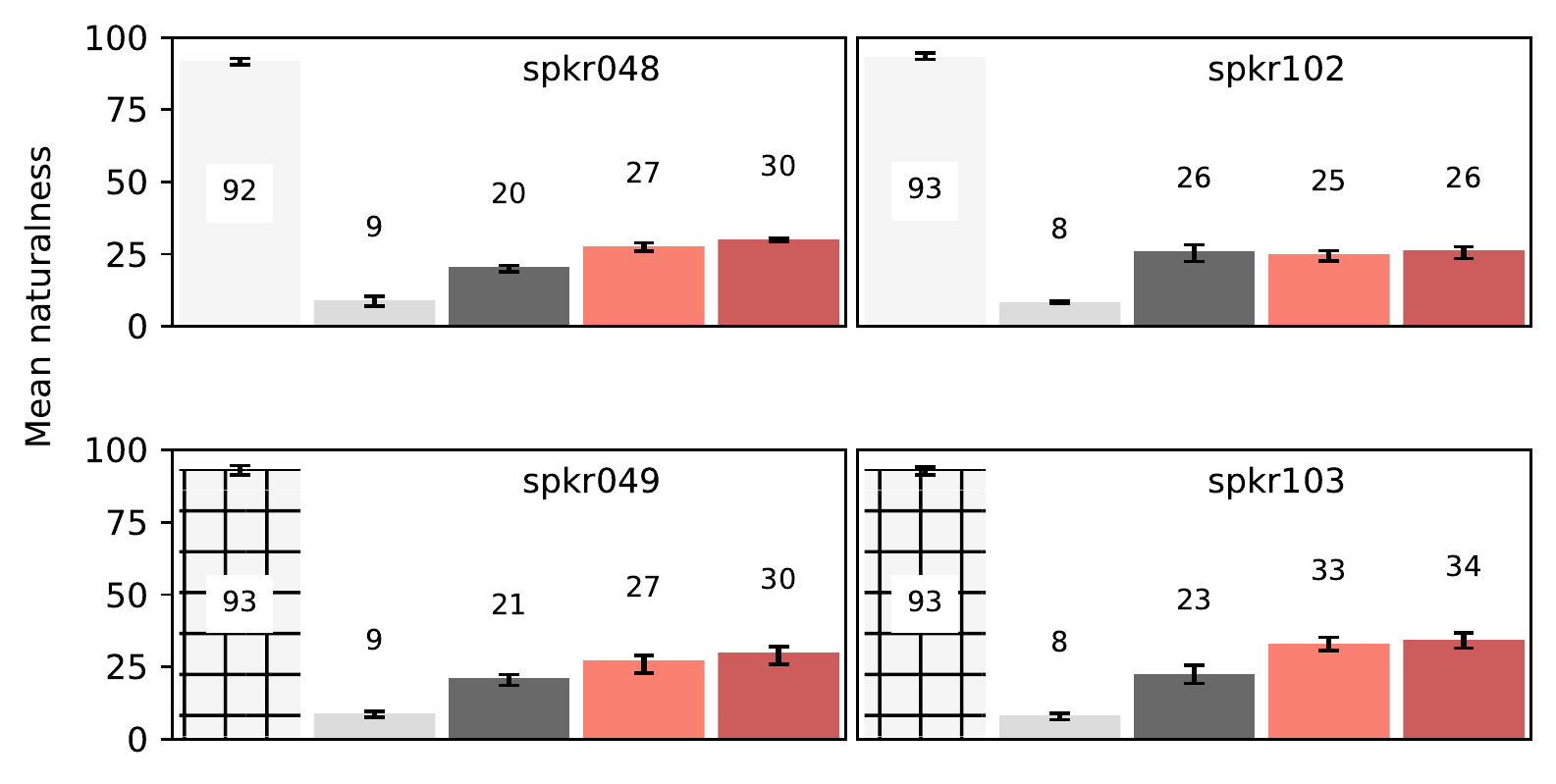}

\vspace{2mm}

\includegraphics[trim=0.25cm 0.3cm 0.2cm 0.05cm, clip=true, width=\columnwidth]{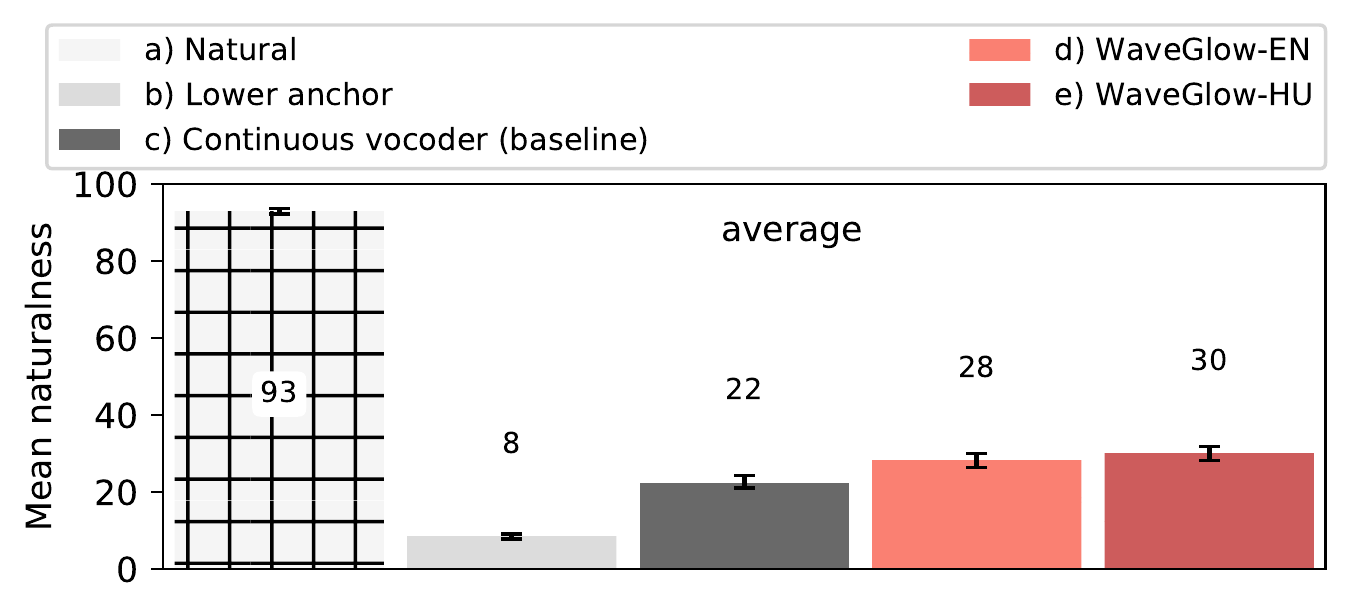}
\caption{\textit{Results of the subjective evaluation with respect to naturalness, speaker by speaker (top) and average (bottom). The errorbars show the 95\% confidence intervals.}}
\vspace{-2mm}
\label{fig:results_subjective}
\end{figure}

Each sentence was rated by 22 native Hungarian speakers (24~females, 2 males; 19--43 years old). On average, the test took 13 minutes to complete. Fig.~\ref{fig:results_subjective} shows the average naturalness scores for the tested approaches. The lower anchor version achieved the lowest scores, while the natural sentences were rated the highest, as expected. The proposed neural vocoder based versions (WaveGlow-EN and WaveGlow-HU) were preferred over the baseline continuous vocoder, except for speaker \#102, for whom they were rated as equal. In all cases, WaveGlow-HU was slightly preferred over WaveGlow-EN.

To check the statistical significances we conducted Mann-Whitney-Wilcoxon ranksum tests with a 95\% confidence level. Based on this, both WaveGlow-EN and WaveGlow-HU are significantly different from the baseline vocoder, but the difference between the English and Hungarian WaveGlow versions is not statistically significant. When checking the significances speaker by speaker, the same tendencies can be seen (WaveGlow-EN $=$ WaveGlow-HU $>$ baseline), except for speaker \#102, for whom the differences are not statistically significant between the baseline and proposed systems.

As a summary of the listening test, a significant preference towards the proposed WaveGlow-EN/WaveGlow-HU models could be observed.

\section{Discussion and conclusions}

In this paper, we used speaker-dependent convolutional neural networks to predict mel-spectrogram parameters from ultrasound tongue image input (in raw scanline representation). The synthesized speech was achieved using WaveGlow inference (trained separately with English and Hungarian data). We compared the proposed model with a baseline continuous vocoder, in which continuous F0, Maximum Voiced Frequency and MGC-LSP spectral features were predicted separately. 

The results of the objective evaluation demonstrated that during the articulatory-to-acoustic mapping experiments, the spectral features are predicted with lower Mel-Cepstral Distortion using the proposed WaveGlow/mel-spectrogram model than with the baseline (5.37~dB vs.\ 5.47~dB). According to the subjective listening test, the WaveGlow flow-based neural vocoder produces more natural synthesized speech compared to the continuous vocoder baseline, for three out of the four speakers. By informally listening to the synthesized sentences of the male speakers, we found that the 80-dimensional mel-spectrogram representation of WaveGlow was not enough to capture changes in F0. Therefore, the final synthesized sentences for male speakers (especially for \#102) are less natural. This could be improved by either a higher dimensional spectral representation (which, in practice is not easy, as both WaveGlow models were trained with 80-D mel-spectrogram), or by non-linearly reshaping the mel-spectrogram during the articulatory-to-acoustic mapping, i.e.\ adding more emphasis to the lower part of the spectrum, which contains F0 information.

The advantage of WaveGlow is that F0 is included in the mel-spectrogram representation, and it is not necessary to predict the excitation separately. In the baseline continuous vocoder~\cite{Csapo2019}, separate CNN models were used for the excitation and spectral prediction. Although here we did not measure the accuracy of F0 prediction separately, from the subjective listening test it is clear that the mel-spectrogram can represent the excitation information well for high F0, but not for low F0 speakers.
The disadvantage of WaveGlow is that for training the neural vocoders, a huge amount of speech data is necessary (24 hours were used for the English model~\cite{Prenger2019} and 22 hours for the Hungarian model). From this point of view, the continuous vocoder is much simpler, as it has 2-dimensional excitation, 25-dimensional spectral features, and no data is required to train the vocoder itself. Besides, it gives controllability, which usually is not fully supported by neural vocoders.

In the future, we plan to apply the above articulatory-to-acoustic prediction framework with the flow-based neural vocoder for other articulatory modalities (e.g.~lip or rtMRI).

The keras implementations are accessible at \url{https://github.com/BME-SmartLab/UTI-to-STFT/}.

\section{Acknowledgements}

The authors were partly funded by the NRDIO of Hungary (FK 124584, PD 127915 grants). T.G.Cs.~and Cs.Z.~were also supported by the European Union's Horizon 2020 research and innovation programme under grant agreement No.~825619 (AI4EU). G.G.~and L.T.~were funded by the J\'anos Bolyai Scholarship of the Hungarian Academy of Sciences, and by the Hungarian Ministry of Innovation and Technology New National Excellence Program ÚNKP-20-5. L.T.~was supported by the TUDFO/47138-1/2019-ITM project. The Titan Xp used for this research was donated by the NVIDIA Corporation. We thank the listeners for participating in the subjective test.

\clearpage

\bibliographystyle{IEEEtran}

\bibliography{ref_collection_csapot_nourl}

\end{document}